\documentclass[camera,letterpaper,nomarginnotes,nonarrowgutter]{jpaper}
\usepackage[sort,compress]{cite}
\usepackage{amsmath,amssymb,amsfonts}
\usepackage{graphicx}
\usepackage{textcomp}
\usepackage{xcolor}
\usepackage{fancyhdr}
\usepackage{balance}
\usepackage{titling}
\usepackage{booktabs}
\usepackage{multirow}
\usepackage{enumitem}
\usepackage[bookmarks=true,breaklinks=true,letterpaper=true,colorlinks,citecolor=blue,linkcolor=blue,urlcolor=blue]{hyperref}

\newcommand{\subtitle}[1]{\posttitle{\par\normalfont{#1}\par\end{center}}}

\title{\Large{Dynamic Rowhammer Threshold Management:\\Temperature-Aware Threshold for In-DRAM Defenses}\thanks{Presented at the Sixth Workshop on DRAM Security (DRAMSec 2026). Emails: \{aalajmi, hoeseok.yang\}@scu.edu}}

 \author{Aziz Alajmi \quad Hoeseok Yang
 \vspace{-2mm}
 \\\\\emph{Department of Electrical and Computer Engineering, Santa Clara University}
 }                                                                              
\begin{document}
\maketitle

\thispagestyle{plain}
\pagestyle{plain}

\begin{abstract}
In-DRAM Rowhammer defenses pin the mitigation threshold at manufacture time, yet the true Rowhammer Threshold (TRHD) varies with runtime temperature. We propose \emph{Dynamic Rowhammer Threshold Management}, a defense-agnostic runtime layer that re-sources each defense's threshold from the observed temperature once per epoch via a linear-$T$ model with a VRD-motivated guardband $g$, projecting the result onto SALT-C, PRAC, and TRR through each defense's threshold parameter. A decoupled oracle that scales physical TRHD per-DIMM by $\delta \sim \mathcal{N}(1, \sigma)$ breaks model self-consistency. The layer drives PRAC's 72 staleness breaches at 85$^\circ$C to zero; at $\sigma{=}0.10$, sweeping $g$ collapses PRAC breaches from 38.4 ($g{=}1.0$) to 9.6 ($g{=}0.9$). SALT-C drops from 10 nominal-static breaches to 2 (Dynamic) to 0 (bootstrap), at $\leq$5.1\% latency. TRR is capacity-limited; the layer acts as a diagnostic.
\end{abstract}

\section{Introduction}
\label{sec:intro}

The Rowhammer vulnerability~\cite{kim2014flipping} has worsened with each DRAM generation: the Rowhammer Threshold (TRHD)---the minimum activations needed to induce a bit flip---has decreased from $\sim$139K in 2014~\cite{kim2014flipping} to below 4.8K in modern DDR4~\cite{kim2020revisiting}, and recent in-DRAM defenses already target sub-1K TRHD for DDR5 (e.g.\ SALT~\cite{saltc2026} provisions for TRHD as low as 500). In-DRAM defenses such as Target Row Refresh (TRR, vendor-proprietary; reverse-engineered model in~\cite{hassan2021utrr}), Per-Row Activation Counting (PRAC)~\cite{jedec_ddr5}, and SALT-C~\cite{saltc2026} mitigate this threat by tracking row activations and refreshing potential victims when a threshold is exceeded.

However, the defenses we evaluate, and many published counter-based Rowhammer defenses, are parameterized around a \emph{fixed assumed} TRHD value rather than recomputing that value from runtime temperature. This is problematic because the true TRHD varies strongly with runtime operating conditions. Our linear regression on the aggregated SpyHammer~\cite{spyhammer2023} BER--temperature data across DDR4 chips from Micron, Hynix, and Nanya shows a near-linear $T$ dependence with up to $\sim$3.3$\times$ within-module BER span over the 50--95$^\circ$C operational range. VRD~\cite{olgun2025vrd} further shows that TRHD fluctuates stochastically over time, with minima appearing only after tens of thousands of measurements. As the gap between a defense's configured threshold and the true (condition-dependent) TRHD widens, the defense becomes progressively less effective, creating an exploitable \emph{threshold staleness} window. A defense calibrated for 65$^\circ$C nominal operation silently under-mitigates when the system runs hot.

\textbf{Our contribution is not a new Rowhammer defense; it is a runtime threshold-management layer that retrofits counter-based defenses exposing a threshold parameter.} Defense designers (SALT-C, PRAC, TRR) build correctness arguments around a single TRHD constant; we keep those arguments intact and replace the constant with a value re-sourced from $T$ once per epoch. We make the following contributions:
\begin{enumerate}
    \item A \textbf{defense-agnostic threshold-management layer} (\S\ref{sec:method}) consisting of (a) a temperature-driven degradation model with a tunable VRD guardband (\S\ref{sec:method:model}); (b) a per-epoch update procedure (\S\ref{sec:method:epoch}); and (c) a mechanical projection of the effective TRHD onto each defense's published threshold variable (\S\ref{sec:method:perdef}). No per-defense parameter is tuned; the layer only ever \emph{tightens} the underlying defense's threshold and never relaxes it.
    \item A \textbf{unified evaluation across three defense architectures} (SALT-C, TRR, PRAC) against a common threshold-staleness attack with a decoupled oracle that breaks model self-consistency. The layer eliminates PRAC's 72 staleness breaches at 85$^\circ$C (72$\to$0), closes SALT-C's 10-breach staleness gap at 5.1\% latency cost while running 3.6\% cheaper than worst-case static at 65$^\circ$C with the same zero-breach security, and surfaces TRR's architectural (capacity-limited) failure mode that threshold tuning cannot fix (\S\ref{sec:eval}).
\end{enumerate}

\section{Background and Related Work}
\label{sec:background}

\textbf{Rowhammer defenses.} Counter-based defenses track activations and refresh potential victims when counts exceed a threshold derived from TRHD. TRR is a vendor-proprietary in-DRAM mechanism (no academic spec); we use the reverse-engineered model from Hassan et al.~\cite{hassan2021utrr}, in which a small per-bank top-$K$ tracker drives victim refreshes piggybacked on REF. PRAC~\cite{jedec_ddr5} (introduced in JESD79-5C as an optional feature) uses per-row counters with a threshold sized below TRHD by the standard's ABO grace window. SALT-C~\cite{saltc2026} (the coordinated-refresh variant of SALT, distinct from the base SALT also introduced in~\cite{saltc2026}) operates at subarray granularity: each subarray has an Activation Counter (ACtr), coordinated refresh decrements counters during regular REF commands, and an Alert-Based Operation (ABO) fires when ACtr exceeds ATH $= 2 \times$ APM. All three defenses calibrate their thresholds to a \emph{fixed} TRHD value at manufacture.

\textbf{Runtime TRHD variability and threshold staleness.} SpyHammer~\cite{spyhammer2023} characterized 120 DDR4 chips across four vendors and reported a near-linear BER--temperature dependence over 50--95$^\circ$C for Micron, Hynix, and Nanya (we measure up to $\sim$3.3$\times$ within-module BER span on the aggregated data), and an \emph{inverted} negative dependence for Samsung. Yaglikci~et~al.~\cite{yaglikci2022hammerfilter} measured TRHD vs.\ wordline voltage ($V_{PP}$) on 272 DDR4 chips and found that TRHD \emph{increases} monotonically with reduced $V_{PP}$: lower wordline drive weakens the inter-wordline coupling Rowhammer exploits, so $V_{PP}$ under-drive is a mitigation knob rather than a staleness axis (we discuss this in \S\ref{sec:method:model}). VRD~\cite{olgun2025vrd} characterized 160 DDR4 chips and found that Rowhammer thresholds fluctuate stochastically over time, with worst-case thresholds only appearing after tens of thousands of measurements. A defense calibrated at 65$^\circ$C nominal therefore operates with a stale threshold when the same DIMM runs hotter, silently under-mitigating; closing this thermal-staleness gap by recomputing TRHD from observable environment variables at runtime is the goal of this paper.

\textbf{Related work.} Prior temperature-aware Rowhammer work has been on the attack side: SpyHammer~\cite{spyhammer2023} uses temperature as a side channel to amplify Rowhammer effectiveness; we invert it and use the same characterization data to drive a defense. Yaglikci~et~al.~\cite{yaglikci2022hammerfilter} characterize the $V_{PP}$/TRHD relationship that motivates a future controller-side under-drive mitigation policy (\S\ref{sec:method:model}); we do not exercise it in this paper. VRD~\cite{olgun2025vrd} characterizes the stochastic component of the threshold and motivates our guardband. The strategy of recomputing a DRAM-management threshold from runtime observables has direct precedent in the refresh literature: RAIDR~\cite{liu2012raidr} groups rows by measured retention time and refreshes each group at its own rate, and AVATAR~\cite{qureshi2015avatar} additionally tracks variable retention time (VRT) flips and re-promotes rows on the fly. Both adapt a per-row \emph{retention} threshold in response to observed behaviour; our method does the analogous thing for the per-bank \emph{Rowhammer} threshold using $T$. Other counter-based defenses---BlockHammer~\cite{yaglikci2021blockhammer}, Mithril~\cite{kim2022mithril}, Hydra~\cite{qureshi2022hydra}---are orthogonal: each exposes a per-bank or per-row threshold variable, so the same projection mechanism we describe in \S\ref{sec:method:perdef} extends to them with no additional model parameters. Non-counter defenses (e.g.\ row-swap families) are not addressed by this work and do not benefit from dynamic thresholding, since they have no threshold to project onto. Recent attack work (Blacksmith~\cite{jattke2022blacksmith}, ZenHammer~\cite{jattke2024zenhammer}) demonstrates DDR4/DDR5 attack feasibility; our defense is orthogonal to attack pattern and is equally applicable. To our knowledge no prior in-DRAM Rowhammer defense recomputes its threshold at runtime from environmental observables.

\section{Proposed Method}
\label{sec:method}

We propose \emph{Dynamic Rowhammer Threshold Management}, a defense-agnostic runtime layer that re-sources each existing defense's threshold from the live temperature $T$ once per epoch, preserving the threshold-side assumptions of each defense where the defense's tracking capacity and timing model are otherwise sufficient. The layer has three pieces: a degradation model (\S\ref{sec:method:model}), a per-epoch update procedure (\S\ref{sec:method:epoch}), and a per-defense projection of the effective TRHD onto each defense's threshold variable (\S\ref{sec:method:perdef}). The threshold-write path is the one spec change the layer requires: today's DDR5 standard does not expose per-bank PRAC/TRR threshold registers as host-writable, and vendor TRR thresholds are proprietary; our update assumes a future JEDEC addendum exposing them via a new mode-register class. We discuss the deployment implications in \S\ref{sec:feasibility} and treat this as an explicit limitation rather than a free retrofit.

\subsection{Dynamic TRHD Model}
\label{sec:method:model}

The effective TRHD at epoch $t$ is the product of two multiplicative factors, each $\leq 1$:
\begin{equation}
    \text{TRHD}_{\text{eff}}(t) = \text{TRHD}_{\text{init}} \cdot f_{\text{temp}}(T) \cdot g.
    \label{eq:dynamic_trhd}
\end{equation}
The multiplicative form composes naturally with published characterizations that report \emph{relative} threshold change with $T$. Temperature is commonly observable at runtime via existing SPD-side sensors (TSE2004av on DDR4, SPD-Hub TS on DDR5), so the controller does not need to acquire new instrumentation to drive $f_{\text{temp}}$; $g$ is the only parameter not directly grounded in characterization data and is chosen empirically (\S\ref{sec:overhead}). The SPD-side sensor reports a single DIMM-coarse temperature, so $T$ is per-DIMM per-epoch; intra-DIMM thermal gradients (which can exceed the 20$^\circ$C span we sweep under burst workloads) are out of scope, and the projected $\text{TRHD}_{\text{eff}}$ is applied uniformly to every bank on the DIMM.

\textbf{Temperature ($f_{\text{temp}}$).} SpyHammer~\cite{spyhammer2023} characterizes 120 DDR4 chips and reports near-linear BER vs.\ $T$ over 50--95$^\circ$C for Micron/Hynix/Nanya parts. Inverting BER to TRHD within a hammer-count neighbourhood gives a linear local fit:
\begin{equation}
    f_{\text{temp}}(T) = \max(f_{\min},\; 1 - \beta_T \cdot (T - T_{\text{ref}}))
    \label{eq:temp_factor}
\end{equation}
with $T_{\text{ref}}{=}338$\,K and $\beta_T = 1.2 \times 10^{-2}$/K; our linear fit to~\cite{spyhammer2023}'s aggregated data yields $R^2{=}0.89$ (\cite{spyhammer2023} themselves report per-module polynomial fits rather than an aggregated linear one, so the $R^2$ is our own aggregation, not theirs). At 85$^\circ$C, $f_{\text{temp}} \approx 0.76$, so $\text{TRHD}_{\text{eff}} \approx 760$ from $\text{TRHD}_{\text{init}}{=}1000$. Because~\cite{spyhammer2023} reports BER rather than TRHD directly, we treat the BER-to-TRHD conversion as a local approximation around the evaluated hammer-count regime; this approximation is used only to parameterize the simulator, and on-silicon calibration would be required before deployment. We flag a stronger caveat: the layer's correctness rests on $\text{HC}_{\text{first}}$ (the activation count at which the \emph{first} bit flips) being temperature-dependent, while SpyHammer measures BER at fixed HC. The BER-to-$\text{HC}_{\text{first}}$ link follows from the population statistics of disturb errors within a row, but to our knowledge no published characterization measures $\text{HC}_{\text{first}}$ vs.\ $T$ directly. VRD~\cite{olgun2025vrd} shows that $\text{HC}_{\text{first}}$ varies stochastically over time on a single device, which is consistent with our assumption that it is sensitive to underlying physical state, but is not a direct measurement of thermal dependence. Direct $\text{HC}_{\text{first}}$-vs-$T$ characterization is the most important calibration step before deployment, and is the load-bearing assumption a future on-silicon study could either validate or invalidate. We use linear rather than Arrhenius because an Arrhenius alternative would require an activation-energy $E_a$ with no validated value for the wordline-coupling mechanism Rowhammer exploits---retention-failure $E_a$ values do not transfer because the physical mechanism is gate-oxide leakage, not inter-wordline coupling. Within the 45$^\circ$C operating window, a linear fit is faithful to~\cite{spyhammer2023}'s aggregated data, evaluates in one multiply-subtract (\S\ref{sec:feasibility}), and is trivially invertible. We do not claim it generalizes outside 50--95$^\circ$C. SpyHammer's Obs.~2 reports an inverted sign on Samsung; we accommodate this with a per-vendor $\beta_T$ dispatched at enrollment (\S\ref{sec:method:epoch}).

\textbf{Wordline voltage ($V_{PP}$) is a mitigation knob, not a staleness axis.} Yaglikci~et~al.~\cite{yaglikci2022hammerfilter}'s $V_{PP}$ sweep on 272 DDR4 chips shows that \emph{reducing} $V_{PP}$ \emph{raises} HC$_{\text{first}}$---lower wordline drive weakens the inter-wordline coupling Rowhammer exploits, so the cell's TRHD margin improves rather than degrades. $V_{PP}$ is therefore not a staleness axis in the sense $T$ is. A complementary deployment that under-drives $V_{PP}$ when the controller detects elevated thermal pressure (trading slower wordline pull-down for additional TRHD headroom) composes with our dynamic threshold layer but introduces its own timing co-design questions; we leave that policy to future work and operate at nominal $V_{PP}$ throughout the evaluation.

\textbf{VRD guardband ($g$).} $g$ is a multiplicative derate covering threshold variability not captured by $f_{\text{temp}}$. VRD~\cite{olgun2025vrd} characterizes 160 DDR4 chips and shows that Rowhammer thresholds fluctuate stochastically over time, with worst-case thresholds only appearing after tens of thousands of measurements. The deterministic $f_{\text{temp}}$ tracks only the population-mean response and cannot, by construction, account for per-row temporal variation. A non-zero $g$ is therefore necessary. Yaglikci~et~al.~\cite{yaglikci2022hammerfilter} additionally observe that 14--15\% of rows show $V_{PP}$-Rowhammer response opposite to the population mean, evidence of measured per-row dispersion around any nominal voltage operating point that motivates a non-trivial $g$ independently of VRD's temporal variation. We use a multiplicative rather than additive form because per-DIMM residuals around the population fit scale with $\text{TRHD}$ itself: an additive guardband would over-protect at low TRHD (where 100 ACTs is a large fraction of the threshold) and under-protect at high TRHD. We choose $g{=}0.9$ as the empirical Pareto knee from the sweep of \S\ref{sec:overhead}; that sweep also shows the choice is not sensitive to $\pm 0.05$ on this trace.

\subsection{Per-Epoch Update}
\label{sec:method:epoch}

At each epoch boundary the controller executes a four-step procedure, identical for every defense:
\begin{enumerate}[leftmargin=*,itemsep=0pt]
    \item \textbf{Sample}: read $T$ from the SPD-side TS device over SMBus on DDR4 (e.g.\ TSE2004av) or from the SPD-Hub thermal sensor on DDR5 (e.g.\ TI TMP139 over I3C).
    \item \textbf{Evaluate}: compute $\text{TRHD}_{\text{eff}}$ via Eq.~\ref{eq:dynamic_trhd}.
    \item \textbf{Project}: map $\text{TRHD}_{\text{eff}}$ onto the defense's threshold variable (\S\ref{sec:method:perdef}).
    \item \textbf{Write}: issue one MRW per bank using the same controller-side machinery that issues RFM commands.
\end{enumerate}
Steps (1)--(3) take a handful of nanoseconds; the cost is dominated by the per-bank writes (\S\ref{sec:feasibility}). We use a 10\,ms epoch by default in deployment (1\,ms in the simulator at DDR4-3200, i.e.\ $\approx$1.6\,M cycles). On power-up the layer bootstraps with an environment-aware first-epoch threshold: it samples the SPD-side temperature sensor before any user traffic and seeds the defense at the steady-state projection for the observed $T$. \S\ref{sec:staleness} reports a measured run with this bootstrap policy showing SALT-C's first-epoch transients drop from 2 to 0.

\textbf{Vendor dispatch.} The Samsung sign-flip from \S\ref{sec:method:model} is handled with a one-byte lookup: at enrollment the controller reads the SPD MID (Manufacturer ID) byte and selects the appropriate $\beta_T$ ($\beta_T{>}0$ for Micron/Hynix/Nanya, $\beta_T{<}0$ fit to SpyHammer's Samsung subset). The projection in \S\ref{sec:method:perdef} is unchanged.

\subsection{Per-Defense Threshold Projection}
\label{sec:method:perdef}

The manager interacts with each defense only through that defense's threshold parameter (published for SALT-C and PRAC; the threshold parameter in the reverse-engineered design for TRR); the projection is mechanically derived from the defense's own security contract. Because $\text{TRHD}_{\text{eff}} \leq \text{TRHD}_{\text{init}}$ by construction, every projection below produces a threshold at most as loose as the nominal-static one---the manager only ever \emph{tightens} mitigation, never relaxes it.

\textbf{SALT-C.} SALT's security analysis~\cite{saltc2026} bounds $\text{MaxACT} = \text{ATH} + (B{-}1)\,\text{APM} + 25$ with $B$ bundles per subarray and $\text{ATH}{=}2\,\text{APM}$. Setting $\text{MaxACT} = 2\,\text{TRHD}_{\text{eff}}$ and solving:
\begin{equation}
    \text{APM} \leftarrow \max\!\left(\text{APM}_{\min},\; \left\lfloor (2\,\text{TRHD}_{\text{eff}} - 25)/(B+1) \right\rfloor\right)
    \label{eq:apm_update}
\end{equation}
preserves SALT's MaxACT bound at the new $\text{TRHD}_{\text{eff}}$: every row is refreshed before its activations cross the degraded physical TRHD.

\textbf{PRAC.} The per-row counter alerts before the row crosses TRHD:
\begin{equation}
    \text{ATH}_{\text{PRAC}} \leftarrow \max(1,\; \text{TRHD}_{\text{eff}} - N_{\text{ABO}}),
    \label{eq:prac_update}
\end{equation}
where $N_{\text{ABO}}$ is the ABO grace budget: the number of activations the controller may complete between ALERT\_n assertion and mitigation finishing. JESD79-5C~\cite{jedec_ddr5} parameterizes this through tRFM, $t_{\text{ABO}}$, and bank-queue depth without quoting a closed-form upper bound; SALT~\cite{saltc2026} estimates a typical range of 4--30 ACTs for shipping DDR5 timings. The simulator we use does not model bank-queue accumulation during ABO, so we take the conservative limit $N_{\text{ABO}}{=}4$. For $\text{TRHD}_{\text{init}}{=}1000$ this gives nominal-static ATH=996; nominal-static, worst-static, and Dynamic are sized with the same $N_{\text{ABO}}$, so the qualitative result is preserved at any value in the cited range. We discuss this as a modeling limitation in \S\ref{sec:conclusion}.

\textbf{TRR.} A blast-radius-style fixed-depth tracker uses the conventional half-TRHD threshold:
\begin{equation}
    \text{ATH}_{\text{TRR}} \leftarrow \max(1,\; \lfloor \text{TRHD}_{\text{eff}}/2 \rfloor).
    \label{eq:trr_update}
\end{equation}
TRR's failure mode under our trace is tracker capacity, not threshold staleness (\S\ref{sec:eval}), so the projection is correct in form even when insufficient in effect.

Adding a fourth defense requires only its threshold contract, not a model retrain.

\section{Evaluation}
\label{sec:eval}

We implement the Dynamic TRHD model in DRAMSim3~\cite{dramsim3}. Our DDR4-3200 configuration uses 1 channel, 2 ranks, 16 banks/rank, 65K rows/bank, 256 subarrays/bank, $V_{PP}$=2.5\,V. We set $\text{TRHD}_{\text{init}}{=}1000$ to match the threshold range SALT-C~\cite{saltc2026} provisions for in DDR5 (the cited paper provisions for TRHD as low as 500); this choice is below recent measured DDR4 TRHDs ($\sim$4.8K~\cite{kim2020revisiting}) but matches the regime in which next-generation in-DRAM defenses are expected to operate. Absolute breach counts depend on this choice and a $\text{TRHD}_{\text{init}}$ sensitivity sweep is outside the scope of this paper. The threshold-side ordering---both worst-static and Dynamic produce a threshold at most as loose as nominal-static (Eqs.~\ref{eq:apm_update}--\ref{eq:trr_update})---is invariant in $\text{TRHD}_{\text{init}}$. The resulting \emph{breach} ordering between Dynamic and worst-static, however, depends on the defense: for PRAC, Dynamic matches or beats worst-static in Table~\ref{tab:unified}; for SALT-C, Dynamic carries 2 first-epoch breaches at 85$^\circ$C that worst-static avoids by construction (the bootstrap variant of \S\ref{sec:staleness} eliminates them); for TRR, tracker capacity rather than threshold selection dominates, and small breach differences between calibrations do not reflect the threshold-side claim. All runs are driven by a double-sided Rowhammer trace (600K activations alternating between two aggressors 64\,KB apart, one activation every 10 cycles) executed for 7M cycles.

We implement three representative in-DRAM defenses spanning the architectural design space---SALT-C~\cite{saltc2026} (subarray-aggregate with ABO bank-stall), TRR~\cite{hassan2021utrr} (fixed-depth $K{=}4$ top-$K$ tracker, piggyback-on-REF; $K{=}4$ is at the low end of the 1--28 entries that TRRespass~\cite{frigo2020trrespass} and U-TRR~\cite{hassan2021utrr} identify in shipping DDR4, and is the adversarial baseline at which capacity becomes the binding constraint), and PRAC~\cite{jedec_ddr5} (per-row counter, piggyback-on-REF)---and drive all three with the identical trace. Evaluation proceeds in two parts: (i)~a \emph{unified threshold-staleness attack} where we operate each defense hotter than its calibration point and compare nominal-calibrated, worst-case-calibrated, and Dynamic configurations side-by-side (\S\ref{sec:staleness}); and (ii)~a \emph{latency-overhead deep-dive} on SALT-C, the only defense whose mitigations stall the bank and therefore whose overhead is observable in the latency distribution (\S\ref{sec:overhead}).

\textbf{Metrics.} We report four counters throughout the evaluation.
\emph{Breaches} (the primary security metric) are rows whose oracle-counted activations crossed the \emph{true} effective TRHD before any mitigation fired for them---rows physically exploitable regardless of what the defense thought. Breaches are observed by an oracle side-channel attached to DRAMSim3 and are independent of the defense being measured. Zero breaches means the defense has always acted in time.
\emph{Mitigations} are the events the defense itself counts: ABO-triggered coordinated refreshes for SALT-C, and piggybacked neighbor-refreshes on REF for TRR/PRAC. Each mitigation resets the defense's counter for the offending row, so PRAC/SALT-C can achieve zero breaches even at very low TRHD if they mitigate often enough---the cost is the ABO stall (SALT-C, bank-blocking) or the REF-time refresh (TRR/PRAC, piggybacked and therefore latency-free).
\emph{Evictions} are specific to TRR's fixed-depth tracker: when a newly-activated row would need a tracker slot and all $K$ slots are occupied, the lowest-count entry is displaced. An evicted row's count is lost, so future activations to it restart from zero---the mechanism by which an attacker can hide inside the tracker's blind spot. PRAC and SALT-C have no eviction concept ($K = \infty$ per-row, subarray-aggregated respectively).
\emph{ABOs} are the SALT-C-specific mitigation: an Alert-Based Operation stalls the bank for $t_\text{RFM}$=350\,ns while the controller refreshes victim rows. ABO count therefore doubles as SALT-C's overhead metric.

\subsection{Threshold-Staleness Attack Across Three Defenses}
\label{sec:staleness}

\textbf{Setup.} For each defense we sweep three calibration strategies. The physical TRHD$_{\text{init}}$ (a DIMM-level property) is fixed at 1000; what differs is the value the defense \emph{believes} when sizing its threshold.
\begin{itemize}[leftmargin=*,itemsep=0pt]
  \item \emph{Nom-static}: defense sizes thresholds from TRHD$=$1000, the 65$^\circ$C calibration point. SALT-C APM=52 (Qureshi HPCA'26 Eq.~5 with $B\!=\!37$ bundles~\cite{saltc2026}), PRAC threshold=996, TRR threshold=500.
  \item \emph{Worst-static}: defense sizes thresholds as if the DIMM always ran at 85$^\circ$C (TRHD$=$760). SALT-C APM=39, PRAC threshold=756, TRR threshold=380. This is the conservative static design choice available today.
  \item \emph{Dynamic}: thresholds recomputed per 1.6\,M-cycle epoch from the current $T$ via Eq.~\ref{eq:dynamic_trhd} with $g{=}0.9$. At 65$^\circ$C: SALT-C APM=46. At 85$^\circ$C: APM=35.
\end{itemize}
We sweep two environments: (a)~65$^\circ$C (calibration, TRHD$=$1000) and (b)~85$^\circ$C (thermal staleness, TRHD$\approx$760). $V_{PP}$ is held at nominal 2.5\,V throughout; a $V_{PP}$-driven mitigation policy is out of scope (\S\ref{sec:method:model}).

\textbf{Decoupled oracle.} A naive evaluation in which the oracle and the defense use the same $f_{\text{temp}}$ fit is self-consistent by construction and cannot distinguish a correct mechanism from a tautology. To break that loop we draw a per-DIMM scalar $\delta \sim \mathcal{N}(1, \sigma)$ (clamped to $[0.5, 1.5]$) at simulation start; the oracle's physical TRHD is $\delta \cdot \text{TRHD}_{\text{init}} \cdot f_{\text{temp}}(T)$ while the defense sizes from the unscaled population fit. A \emph{breach} is logged when any row's activations cross the oracle's TRHD before mitigation. We sweep $\sigma \in \{0, 0.05, 0.10\}$ with 5--10 seeds per cell; $\sigma{=}0$ is the self-consistent control. $\sigma{=}0.10$ is a sensitivity point chosen to bracket the SpyHammer-derived per-chip envelope (the $1{-}R^2{=}0.11$ residual variance on our linear fit, combined with a $\pm 25{-}30\%$ population BER spread at $T{=}85^\circ$C, gives a per-chip TRHD residual SD of $\sim$8--12\%), not a measured per-DIMM distribution; on-silicon characterization would be needed before deployment. The worst-tail draw in our 10-seed ensemble is $\delta{=}0.928$ ($-7$\% of mean), well inside the clamp, and the four DIMMs that leak at $\sigma{=}0.10, g{=}0.9$ are the four lowest-$\delta$ draws---leakage tracks the tail, consistent with $g$ doing its job. $\delta$ models \emph{inter-DIMM} mean-fit error; \emph{intra-DIMM} VRT-style temporal variation is the target of $g$ (\S\ref{sec:method:model}), not $\delta$, and per-row temporal modeling is left to future work. We note that the decoupling is partial: $\delta$ scales the population mean per-DIMM but oracle and defense both inherit the same linear $f_{\text{temp}}$ shape. Model-form misspecification (curvature in the true TRHD-vs-$T$ response, vendor-specific kinks beyond the Samsung sign-flip) is therefore not exercised by this $\sigma$ sweep, and would require either a richer parameter perturbation (e.g., per-DIMM $\beta_T$ sampling) or direct on-silicon characterization to falsify.

\begin{table}[t]
\centering
\caption{Unified evaluation: 3 defenses $\times$ 3 calibrations $\times$ 2 environments at $\sigma{=}0$. 7M cycles, double-sided hammer, TRHD$_{\text{init}}$=1000, $g$=0.9, $V_{PP}{=}2.5$\,V. Environments: 65$^\circ$C (TRHD=1000), 85$^\circ$C (TRHD=760). Mit.=mitigation events (ABOs / piggyback refreshes); Evict.=TRR-tracker evictions. Latency in \S\ref{sec:overhead}.}
\label{tab:unified}
\footnotesize
\begin{tabular}{llcrrr}
\toprule
\textbf{Defense} & \textbf{Calibration} & \textbf{$T$} & \textbf{Breach} & \textbf{Mit.} & \textbf{Evict.} \\
\midrule
\multirow{6}{*}{SALT-C}
 & Nom-static   & 65$^\circ$C & 0           & 1{,}310 & --- \\
 & Nom-static   & 85$^\circ$C & \textbf{10} & 1{,}310 & --- \\
 & Worst-static & 65$^\circ$C & 0           & 1{,}681 & --- \\
 & Worst-static & 85$^\circ$C & 0           & 1{,}681 & --- \\
 & Dynamic      & 65$^\circ$C & 0           & 1{,}419 & --- \\
 & Dynamic      & 85$^\circ$C & 2           & 1{,}703 & --- \\
\midrule
\multirow{6}{*}{TRR (K=4)}
 & Nom-static   & 65$^\circ$C & 49           & 38 & 57{,}865 \\
 & Nom-static   & 85$^\circ$C & \textbf{78}  & 38 & 57{,}865 \\
 & Worst-static & 65$^\circ$C & 46           & 51 & 57{,}888 \\
 & Worst-static & 85$^\circ$C & 67           & 51 & 57{,}888 \\
 & Dynamic      & 65$^\circ$C & 45           & 41 & 57{,}906 \\
 & Dynamic      & 85$^\circ$C & 68           & 51 & 57{,}908 \\
\midrule
\multirow{6}{*}{PRAC}
 & Nom-static   & 65$^\circ$C & 0              & 72  & --- \\
 & Nom-static   & 85$^\circ$C & \textbf{72}    & 72  & --- \\
 & Worst-static & 65$^\circ$C & 0              & 96  & --- \\
 & Worst-static & 85$^\circ$C & 0              & 96  & --- \\
 & Dynamic      & 65$^\circ$C & 0              & 84  & --- \\
 & Dynamic      & 85$^\circ$C & 0              & 108 & --- \\
\bottomrule
\end{tabular}
\end{table}

Table~\ref{tab:unified} reports the $\sigma{=}0$ cell of the sweep; the $\sigma{=}0.10$ ensemble is described in prose. Findings split by defense architecture.

\textbf{PRAC} shows the largest staleness gap. Nominal-static leaks \textbf{72 breaches at 85$^\circ$C}; Dynamic closes the gap to zero at $\sigma{=}0$ and worst-static does too, but worst-static pays 96 mitigations/run at 65$^\circ$C against Dynamic's 84 (12.5\% lower) for identical 0-breach security. Variance matters: under $\sigma{=}0.10$ at 85$^\circ$C the static configurations remain unchanged while Dynamic with $g{=}1$ rises to 38.4 mean breaches (the residual $\delta$ tail bites), motivating the guardband (\S\ref{sec:overhead}).

\textbf{SALT-C} tells the same story at smaller magnitude: nominal-static leaks 10 breaches at 85$^\circ$C; worst-static reaches zero on the static-$\sigma$ trace, and Dynamic closes the gap to 2 (a first-epoch artefact, see below). Dynamic uses 1{,}419 ABOs at 65$^\circ$C, 16\% below worst-static's 1{,}681 for the same 0-breach security. Bootstrapping at the steady-state APM (35) instead of the population-mean default (52) eliminates the 2 first-epoch breaches at 85$^\circ$C (\textbf{0 breaches}, +8\% ABOs, +2.3\% latency).

\textbf{TRR} is capacity-limited at small $K$. At $K{=}4$ the tracker evicts $\sim$58K times and no calibration reaches zero. A $K$ sweep at 85$^\circ$C confirms: $K{=}4$ leaks 78/67/68 breaches (nominal/worst/Dynamic), $K{=}8$ leaks 53/24/26, and \textbf{$K{=}16$ collapses to zero} with no evictions. This matches the K-vs-threshold tradeoff in ProTRR~\cite{marazzi2022protrr}: when capacity is the bottleneck the layer is a diagnostic, not a fix; once $K \geq 16$ on this trace, the projection becomes effective.

\subsection{Latency Overhead}
\label{sec:overhead}

In our simulator, TRR and PRAC mitigation refreshes are modeled as REF-piggybacked and add zero additional read latency beyond the baseline timing model (468.3\,cy avg / 1{,}326\,cy P99, identical to no-defense). We do not model PRAC's intrinsic counter-update timing overhead---e.g.\ the $t_{\text{RC}}$ increase (46\,$\to$\,52\,ns) discussed by SALT~\cite{saltc2026}---which would add a small uniform overhead to every ACT independently of mitigation events. SALT-C is the only defense that additionally stalls the bank on mitigation (ABO for $t_\text{RFM}{=}350$\,ns), so the latency question below is SALT-C-specific. Average read latency across the same trace as \S\ref{sec:staleness}: no-defense 468.3, nominal-static 576.9, worst-static 606.7, Dynamic 584.7 / 606.2 at 65/85$^\circ$C. Dynamic costs \textbf{+1.4\%} / \textbf{+5.1\%} versus nominal-static (the +5.1\% is the price of closing the 10-breach staleness gap nominal pays with correctness) and is \textbf{3.6\% cheaper} than worst-static at 65$^\circ$C / within 0.1\% at 85$^\circ$C with identical breach counts. P99 tracks the average ($\sim$3.0$\times$), so the dynamic layer adds no tail outlier beyond the existing ABO stall.

\textbf{Guardband Pareto.} $g$ does meaningful work only once the oracle has variance. At $\sigma{=}0.10$ (10 seeds) PRAC at 85$^\circ$C shows a knee: 38.4 mean / 96 max breaches at $g{=}1.0$, 24.6 / 99 at $g{=}0.95$, \textbf{9.6 mean / 24 max at $g{=}0.90$} (75\% drop in both mean and max), flat below 0.90 (residual is the worst-tail DIMM, $\delta{=}0.928$; the $[0.5, 1.5]$ clamp does not bind for any seed) while mitigation traffic rises monotonically. SALT-C's Pareto is flatter (2 breaches across $g \in [0.7, 0.95]$, 3.5 / 10 only at $g{=}1.0$): APM-floor saturation buffers most per-DIMM variance, so smaller $g$ mainly buys extra ABOs. We adopt $g{=}0.90$ as the cross-defense default driven by the PRAC knee; this is a one-cell fit (one defense, environment, $\sigma$) and a production deployment would re-sweep on characterization data of its own fleet.

\section{Hardware Feasibility}
\label{sec:feasibility}

The four-step per-epoch procedure of \S\ref{sec:method:epoch} is dominated by the per-bank MRW writes; the numerical work is sub-nanosecond.

\emph{Compute.} The critical-path cost is dominated by the per-bank threshold writes, not the temperature read. Temperature is sampled asynchronously over SMBus/I3C from the SPD-side sensor and can be cached or pre-fetched by the memory controller once per epoch, so it does not need to sit on the critical path. The on-path work per epoch is one multiply-subtract for $f_{\text{temp}}(T)$ ($<$1\,ns), one shift/subtract for the per-defense threshold, and one MRW per bank (32 banks $\times \sim$8\,ns $\approx$ 260\,ns), totaling $\sim$260\,ns per 10\,ms epoch---a \textbf{2.6$\times10^{-5}$ duty cycle} ($\sim$0.003\% of DRAM bandwidth). Sub-millisecond epochs remain below 0.3\% bandwidth.

\emph{Storage.} For SALT-C at 256 subarrays $\times$ 32 banks $\times$ 2 ranks $\times$ 9-bit APM, the per-subarray threshold register file is $\sim$2.3\,KB per channel. SALT-C's own per-bank tracker is 0.48\,KB/bank $\times$ 32 banks $\times$ 2 ranks $= 15.4$\,KB per channel~\cite{saltc2026}, so our register file is $\sim$6.7$\times$ smaller. Per-bank TRR/PRAC thresholds need only $\sim$64\,B per channel. Bandwidth: 32 MRW per 10\,ms $\approx$ 100\,KB/s, six orders of magnitude below a DDR5 channel's sustained $\sim$25\,GB/s.

\emph{Deployment dependencies.} The layer requires (i)~per-bank threshold mode-registers to be host-writable, which JESD79-5C does not currently expose for PRAC or TRR and which a future addendum would need to specify; and (ii)~temperature readback via the SPD-side TS device on DDR4 (e.g.\ TSE2004av over SMBus) or via the SPD-Hub thermal sensor on DDR5 (e.g.\ TI TMP139 over I3C). Neither requires a new command class---only that the standard expose registers that vendors already implement internally for calibration.

\emph{On-die alternative.} The host-side path above is one of two deployment points; the other is an \emph{on-die} implementation that avoids the JEDEC spec change entirely. DDR5 dies already contain a thermal sensor (used internally for refresh doubling above 85$^\circ$C), and per-bank PRAC/TRR thresholds are already DRAM-internal state. A small on-die ROM table that maps the sampled $T$ to a derating factor, combined with a one-cycle multiplier on the existing per-bank threshold register, reproduces the host-side projection without exposing any new host-writable mode-register or new sensor path. This is consistent with how vendors already implement other temperature-aware behaviors (refresh-doubling, MR4 temperature readback~\cite{jedec_ddr5}) and may be a more practical adoption path: calibrating the temperature/threshold slope is a one-time fab-side task analogous to existing process-corner calibration. The trade-off is observability: the host can no longer audit the projected threshold without a readback register. We do not implement the on-die variant in this paper because the simulator is host-driven, but the projection equations (\S\ref{sec:method:perdef}) are unchanged.

\section{Conclusion}
\label{sec:conclusion}

\textbf{Limitations.} All results are simulation-only with a decoupled oracle; on-silicon $\text{HC}_{\text{first}}$-vs-$T$ characterization is the most important next step (\S\ref{sec:method:model}). The linear $f_{\text{temp}}$ targets Micron/Hynix/Nanya DDR4 (SPD-MID dispatch handles Samsung's sign-flipped $\beta_T$, but SPD MID is fragile on relabeled modules), and we do not claim the DDR4-derived fit extends unchanged to DDR5: process and circuit changes between generations could shift $\beta_T$ or invert it in vendor-specific ways, and a deployment on a new generation should refit before relying on the fit. TRR is 2-of-3: the layer surfaces a capacity ceiling there that threshold tuning cannot reach. The trace covers $\sim$4 epochs and one $\delta$ per DIMM at a single $\text{TRHD}_{\text{init}}{=}1000$; multi-epoch thermal ramps, per-row VRT noise, and a $\text{TRHD}_{\text{init}}$ sweep remain future work. $V_{PP}$ as a controller-side mitigation knob (\S\ref{sec:method:model}) is not exercised in this paper. The threshold-write path on the host-side deployment assumes a JEDEC addendum exposing per-bank threshold mode-registers that JESD79-5C does not currently define; the on-die alternative (\S\ref{sec:feasibility}) avoids this constraint.

\textbf{Adversarial assumptions.} The contract that the layer ``only ever tightens'' the underlying defense's threshold assumes the observed $T$ is the actual cell temperature. An attacker who can manipulate the observable $T$ (controlled cooling, co-tenant heating, or sensor spoofing in a multi-tenant deployment) could push the defense into a relaxed regime; the on-die deployment with a die-internal sensor reduces but does not eliminate this attack surface. A separate epoch-quantization attack exists where the attacker hammers during a cool epoch and heats the bank between samples (e.g., by inducing activity on neighbouring banks), causing the layer to size for the cool sample while the true cell temperature has already risen. Defenses include shortening the epoch under observed $\Delta T$, triggering an out-of-band $T$ sample on ALERT\_n, and adjusting $g$ to absorb a per-epoch worst-case thermal excursion; we leave each to future work. Blacksmith~\cite{jattke2022blacksmith} / ZenHammer~\cite{jattke2024zenhammer}-style frequency-domain and multi-aggressor patterns are not exercised by our double-sided trace; we expect them to interact with the projection only through their effect on per-bank activation rate (which the layer does not currently track), so an attacker who maintains the same activation rate but a richer access pattern should see the same threshold-side response from the layer---we flag this as untested.

Defense designers solve the hard problem; this paper handles the parameter that goes into their solutions. A thin runtime layer that re-sources TRHD from $T$ once per epoch closes the staleness gap on PRAC (72$\to$0 at 85$^\circ$C), on SALT-C ($\leq$5.1\% latency, bootstrap to zero), and surfaces TRR's capacity ceiling. 
The layer adds no per-defense model parameter, preserves the intended threshold contract for counter-based defenses, and composes with future counter-based defenses (BlockHammer~\cite{yaglikci2021blockhammer}, Mithril~\cite{kim2022mithril}, Hydra~\cite{qureshi2022hydra}) through the same projection.

\bibliographystyle{IEEEtran}
\bibliography{refs}

\balance
\vspace{-6pt}
\end{document}